\documentclass[twocolumn,showpacs,preprintnumbers,amsmath,amssymb]{revtex4}

\begin{document}
\draft

\title{Quantum capacitance: a microscopic derivation}

\author{Sreemoyee Mukherjee$^1$,
M. Manninen$^2$ and P. Singha Deo$^1$}
\address{$^1$Unit for Nano Science and Technology,
S. N. Bose National Centre for Basic Sciences, JD Block,
Sector III, Salt Lake City, Kolkata 98, India.}
\address{$^2$Nano-science Center, University of Jyvaskyla, PO Box-35,
40100 Jyvaskyla, Finland.}
\date{\today}

\begin{abstract}
We start from microscopic approach to many body physics and show
the analytical steps and
approximations required to arrive at the concept of quantum
capacitance.  These approximations
are valid only in the semi-classical limit and the quantum capacitance
in that case is determined by Lindhard function. The effective
capacitance is the geometrical capacitance and the quantum capacitance
in series, and this too is established starting from a microscopic
theory. 
\end{abstract}


\maketitle

\section{Introduction}

Several new concepts and ideas have developed in last few decades
on nano-electronics and they are often questioned \cite{das}.
AC response of quantum dots in the coherent regime
has been measured in recent experiments \cite{gab,fev,kat,blu,kae}. 
A good understanding and control over such phenomenon
can lead to many novel devices, specially in metrology \cite{gab,nig2}.
The experimental results are
analyzed in a series of works, using effective variables like quantum
capacitance \cite{nig2,mos1,mos2,nig1,nig3,nig4}. 
Capacitance of mesoscopic systems are very different from geometric
capacitance. In mesoscopic systems 
one can differentiate between electrostatic
capacitance and electro-chemical capacitance. Although, in principle
one can also define a magnetic field induced 
capacitance, in practice one defines
a field dependent electrostatic or electrochemical capacitance \cite{scr}.
Ref \cite{but93} gives a detailed analysis of electrochemical 
capacitance which gives corrections to the geometrical capacitance due
to field penetration into the conductor which occurs over a finite
length scale comparable to the dimensions of the sample and ignored
in large systems. Electrochemical capacitance is a property of
open systems (systems connected to leads and electron reservoirs).
In such open systems, electron-electron interaction cannot be treated
exactly and characteristic potentials were introduced
to account for Coulomb interaction approximately.
The correction term appear as another capacitance in series with the
geometric capacitance. Both, open and
closed systems can have an electro-static
voltage induced capacitance. A closed system of a finite wire (referred
as a stub)
connected to a closed ring was analyzed in Ref. \cite{scr}.
The system was reduced to a two level system, wherein there is
a hybridization of a single level coming from the stub and another
single level coming from the ring. Coulomb interaction was again
treated approximately with the help of characteristic potentials
and single particle level approximations. Quantum corrections 
was again shown to appear as a capacitance that appear in series
with the geometric capacitance.  The quantum capacitance
is given by the Lindhard function \cite{scr}.
Subsequently, several authors have tried to interpret experimental
data and numerical calculations in terms of quantum capacitance
\cite{shy,pol}.
A microscopic analysis stating under what circumstances and
assumptions one can use such a parameter is not done so far.

Capacitance of a system is self consistently determined
by Coulomb interactions and this is no exception for quantum
capacitance as well. However, quantum mechanically electrons
can also interact via Fermi statistics and so even when Coulomb
interaction is ignored, a system can have a quantum capacitance.
While geometric or classical capacitance is determined by the volume,
shape and dielectric constant of the system, charge in quantum
mechanics can reside in orbitals that do not have a space-time
description.  The existence of an
effective variable of quantum capacitance,
can simplify the complexity of many body physics.

Unlike that in open systems, electron-electron interaction and 
statistics can be treated exactly in closed systems.
In this work we deal with closed systems so that an
analytical proof can be given and exact numerical
diagonalization is possible for verification.
We would like to analyze the assumptions and concepts 
required to arrive at a statistical mechanical variable
of electrostatic quantum capacitance.
When a system is weakly coupled to
a reservoir, making it an open system, one can describe the
system in terms of the eigenenergies of the system and Fermi-Dirac
distribution function \cite{but85}.
So our results are also valid for weakly coupled open systems.
Electro-chemical potential also works by affecting the electrostatic
potential inside the system \cite{but93}. 
So if electrostatic quantum capacitance
cannot be proved then electro-chemical quantum capacitance may
also not hold.

Our analysis is independent of model and valid for any arbitrary
geometry in any dimension.
However, we will use some models and systems for numerical verification,
that are described in section II. 
Analytical derivation of quantum capacitance is given in section III.
Conclusions are given in section IV.

\section{models for numerical verification and illustration}

We have given in Fig. 1, schematic diagrams of a ring (Fig. 1a),
a stub connected to a ring (Fig. 1b), and a 2D square geometry
(Fig. 1c). Although our analysis is not restricted to these
geometries, we will use them as reference
and examples. Figs. (1a), (1b) and (1c) represents continuum cases
whereas Figs. (1d), (1e) and (1f) represent discrete versions of the
same systems as that in (1a), (1b) and (1c), respectively. Discrete
models are useful for numerical analysis.
The vector potential due to a magnetic field
can non-trivially change the electronic states of a system (due
to quantum interference) while
having very little effect on the bound positive charges that
can be assumed to be uniform \cite{scr}.
It is very easy to see polarization due to vector potential
in rings as the magnetic field can remain
confined to the center of the ring while the electrons in the ring
feel only the vector potential. However, it also occurs in small
two-dimensional or three-dimensional quantum systems where weak
magnetic fields have negligible effect (Lorentz's
force being weighted down by the velocity of light), but vector
potential will drastically change the states of the system due
to interference effects. For numerical verification,
we use the generalized Hubbard Hamiltonian describing a discrete system
consisting of sites.
$$H=\sum_{i,\sigma} \epsilon_i C_{i,\sigma}^\dagger C_{i,\sigma} 
+\sum_{<ij>,\sigma}(tC_{i,\sigma}^\dagger C_{j,\sigma} +HC)
+\sum_i U_1 n_{i,\uparrow}n_{i,\downarrow}$$
\begin{equation}
+\sum_{<ij>,\sigma,\sigma'}
U_2 n_{i,\sigma} n_{j,\sigma'}
\end{equation}
where $\epsilon_i$ is the site energy of the $i$th site, $t$ is the
hopping parameter (in presence of magnetic field it becomes
complex, i.e., $t \rightarrow texp[i{\phi \over N \phi_0}$]), 
$U_1$ and $U_2$ are respectively, the
on site and nearest neighbor Coulomb interaction.
In Fig. 2 we show a three dimensional ring (shaded region)
with a flux $\phi$ through the
center of the ring that can cause polarization. 
At a particular point $r$ on the ring we can
bring an STM tip at a voltage $V$ to cause further polarization at that
position $r$ (or $i$) while another STM tip can measure the
local potential $V(r)$ (or $V_i$) at $r$ (or $i$).  
The polarization charge in a segment of the ring can be measured
by a cylinder around the ring by looking at the induced charge
on this cylinder (unshaded contour in Fig. 2). 

\section{analytical derivation}

We will outline here all the mathematical steps required to
describe the polarization of a quantum system
in terms of electrostatic quantum capacitance. When
assumptions are used, we will give numerical verification
and also cite appropriate earlier works.
Suppose the potential $V(r)$ at a point $r$ is changed infinitesimally
giving rise to a delta potential term in the Hamiltonian,
of the form $dV^{ext}(r)= \sum_n\kappa \delta(r-r_n)$, where $r_n$ is the
coordinate of the $n$th electron. 
The Kohn-Hohenberg theorem \cite{dft} states that the energy is an unique
functional of the local potential. But we just use first order perturbation
correction to the energy which is in terms of the applied potential
only and no self consistency is required.
So the increase in energy of the system can be expanded as
$$\Delta E= \sum_n\int d^3r_1 d^3r_2...d^3r_M \psi^*(r_1, r_2, ... r_M) 
\kappa \delta(r-r_n)$$
\begin{equation}
\psi(r_1,r_2, ...., r_M) + O(\kappa^2) + O(\kappa^3) +....
\end{equation}
As $\kappa \rightarrow 0$, then
\begin{equation}
{\partial E \over \partial V(r)}=Q(r)
\end{equation}
where
\begin{equation}
Q(r)=M\int d^3r_2...d^3r_M \psi^*(r, r_2, ... r_M) \psi(r,r_2, ...., r_M) 
\end{equation}
is the charge at $r$.
We verify this numerically for all the geometries shown in Fig. 1.
A plot is shown in Fig. 3 for a disordered ring whose site energies
$\epsilon_i$ vary from $-0.5t$ to $0.5t$.
Other parameters are explained in figure caption.
We have used a single disorder configuration as the
agreement is equally same for all other configurations.
The figure shows the correctness of Eq. 3.
We stress that we use exact diagonalization using Lanczos algorithm
to determine $E$ and
$Q_i$ and hence this is a numerical verification of Eq. 3.

Therefore, we can define a Lindhard function $\eta(r)$.
\begin{equation}
\eta(r)=-{\partial Q(r) \over \partial V(r)}=
-{\partial^2E \over \partial V(r)^2}
\end{equation}
The last step follows from Eq. 3.
For the geometry in Fig. 1(a) or 1(b) or 1(d) or 1(e),
we can change the magnetic flux $\phi$
through the center. This will
immediately cause a redistribution of electronic charge in
every site of the system.  This is shown in Fig. 4 for the
site numbered 8 in a 11 site ring.
There is no qualitative difference
between Fig. 1(d) and 1(e)
even when the stub is weakly coupled to the ring.
Also consider a case when
an external voltage $dV_i^{ext}$ is applied at site $i$
(in the discrete model one can change $\epsilon_i$ infinitesimally
as $V_i$ and $\epsilon_i$ are linearly related).
Obviously one can change both, $\phi$ and $V_i^{ext}$ simultaneously.
Charge at site $i$, $Q_i$ will change.
Due to electron-electron interaction the local potential
$V_i$, will change. Such a potential
change will in turn have a feedback effect on charge displacement
to determine $dV_i$.
This feedback effect is a purely quantum effect as this feedback
occurs because the initial charge displacement (in this case induced
by externally changed flux or potential) can change the quantum
states or eigen-energies of the system. The externally applied
flux and potential also work indirectly by affecting the
quantum states of the system. In other words external changes
give rise to electron displacements for which potential at site 
$i$ (or $j$) change, and this in turn gives
rise to further charge displacements that is self consistently
determined by Coulomb interactions and Fermi statistics.
$dV_i$ (or $dV_j$) is this self consistently determined increment
in potential at site $i$ (or site $j$) and $dQ_i$ is the self consistently
determined charge at site $i$.
We can begin by writing for the discrete system
\begin{equation}
dQ_i={\partial Q_i \over \partial \phi} d \phi +
{\partial Q_i \over \partial V_i} d V_i +
\Sigma_{j\ne i} {\partial Q_i \over \partial V_j} dV_j
\end{equation}
Although $Q_i$ is a functional of $V_i$, one can change the potential
infinitesimally at a particular point without changing the potential
at any other point to define a partial derivative and this is
in fact done to arrive at the concept of functional derivative \cite{com}.
Not to mention, that in the standard definition for total derivatives,
$\partial Q_i \over \partial V_i$ is change in
$Q_i$ due to an infinitesimal test change in $V_i$ (i.e., 
$V_i^{ext}$), but $dQ_i$ on
the LHS and $dV_i$ and $dV_j$ on the RHS are actual changes
which in these systems are determined self consistently.
By the virtue of the fact that we are considering a sum over $j$
makes our treatment valid for any geometry and any dimension, 
where the sum over $j$
will run over all the sites making the system.
For the continuous system Eq. 6 becomes, where the sum is replaced
by an integration,
\begin{equation}
dQ(r)={\partial Q(r) \over \partial \phi} d \phi +
{\partial Q(r) \over \partial V(r)} d V(r) +
\int_{r'\ne r} {\partial Q(r) \over \partial V(r')} dV(r') d^3r'
\end{equation}
Here partial derivative with respect to $V(r)$ means we are changing
the potential in the region $r$ to $r+dr$ infinitesimally \cite{com}.
In the following we will argue analytically and numerically that
\begin{equation}
-\sum_{j\ne i} {\partial Q_i \over \partial V_j}
\approx -\sum_{j\ne i} {\delta Q_i \over \delta V_{j}}
\end{equation}
Or by replacing the sum by integration,
\begin{equation}
\int_{r'\ne r}{\partial \over \partial V(r')}d^3r' \approx \int_{r'\ne r} 
{\delta \over \delta V(r')}d^3r'
\end{equation}
While LHS is a sum of partial derivatives,
the RHS is a functional derivative. 
Integration or sum of all (but one point) partial derivatives on the LHS
is approximately a total derivative. Had this one point ($r=r'$) 
been included
it would have been an exact total derivative with respect to energy
(electronic charge times potential being energy) \cite{pra}. 
So in the above approximation an energy derivative is being replaced
by a functional derivative with respect to local potential.
This approximation is known in other context like deriving the semi-classical
limit of Friedel sum rule \cite{pra}. That means this approximation
is valid in the semi classical regime.
Numerical verification of the approximate equality in Eq. 8 above is shown
in Fig. 5. By doing exact numerical diagonalization it is difficult
to go to a truly semi-classical limit as that will require us to
take a large ring. However, the approximation in Eqs. 8 and 9 are
known in context of deriving the semi-classical limit of Friedel
sum rule \cite{pra}.
If this approximation
holds then one can relate induced voltage
and polarization charge through quantum capacitance as shown below.
Therefore, from Eq. 6 and 8,
\begin{equation}
dQ_i \approx {\partial Q_i \over \partial \phi} d \phi +
{\partial Q_i \over  \partial V_i} d V_i +
\sum_{j\ne i} {\delta Q_i \over \delta V_{j}} dV_{j}
\end{equation}
Similarly, from Eq. 7 and 9,
\begin{equation}
dQ(r)\approx {\partial Q(r) \over \partial \phi} d \phi +
{\partial Q(r) \over \partial V(r)} d V(r) +
\int_{r'} {\delta Q(r) \over \delta V(r')} dV(r') d^3r'
\end{equation}
We have to assume that $dV_j$ is independent of $j$ as further
explained below. Now it follows from charge conservation that
\begin{equation}
-\eta={\partial Q_i \over \partial V_i}=-\sum_{j\ne i}
{\delta Q_i \over \delta V_{j}}
\end{equation}
The RHS is the net change in $Q_i$ due to an infinitesimal functional
increase (or decrease) in the potential at all points except at $i$. 
That is equivalent to not changing the potential anywhere but 
decreasing (or increasing) the potential at $i$ infinitesimally.
Due to charge conservation the change in charge at $i$ will be the same
in both cases.

Coulomb repulsion tends to distribute charge uniformly in a system.
On the other hand quantum interference effect tends to give rise
to un-even distribution of charge.
Assuming that in the semi-classical regime, Coulomb interactions dominate
over quantum interference effects and distributes the charge uniformly,
$dV_j$ becomes independent of $j$ and we denote it as $dV_{rest}$.
Numerical calculations for small size quantum systems show that
for a wide range of parameter space the charge distribution is
uniform \cite{mat, tap}. Only in very low density regime, quantum interference
effects dominate and the charge density breaks up into crests and
troughs \cite{mat, tap}. 
Any measurement process may not be able to resolve
these crests and troughs and may show an average value for the local
potential implying $dV_j$ can be taken to be independent of $j$.
Transverse variation can be mapped to an effective variation
in the longitudinal direction \cite{verg}.
So one can write to a linear order,
\begin{equation}
dQ_i=C(dV_i-dV_{rest})
\end{equation}
where C is the definition of geometric or classical capacitance.
We know that when we ignore quantum interference effects (i.e., large
systems without boundary and impurity effects averaged out or treated
in random phase approximation) we can always get such a linear regime.
Substituting Eq. 12 and Eq. 13 in Eq. 10 and simplifying one gets that

\begin{equation}
(C + {\eta}){(dV_i - dV_{rest}) \over d{\phi}} = {{\partial}Q_i 
\over {{\partial}{\phi}}} = {{\partial}Q_i \over {{\partial}V_{rest}}}
{{\partial}V_{rest} \over {{\partial}{\phi}}} 
\end{equation}
Now since $ V_{rest} $ can be changed by changing $V_i$ or $\phi$,
one can write $ V_{rest}({\phi}, V_i ) $.
Therefore, $$ dV_{rest} = {{\partial}V_{rest} \over {{\partial}{\phi}}}d{\phi} + {{\partial}V_{rest} \over {{\partial}V_i}}dV_i $$
Since, the region indexed $ i $ is very small compared to the rest of the
system, $ {{\partial}V_{rest} \over {{\partial}V_i}}{\rightarrow}0 $
Therefore,
\begin{equation}
{dV_{rest} \over d{\phi}} = {{\partial}V_{rest} \over {{\partial}{\phi}}}.
\end{equation}
From Eq. (14) it follows that
$$ (C + {\eta}){(dV_i - dV_{rest}) \over d{\phi}} = {\eta}{{\partial}V_{rest} \over {{\partial}{\phi}}} = {\eta}{dV_{rest} \over d{\phi}} $$
Multiplying both sides of the above equation by $ C $ we get,
$$ C{(dV_i - dV_{rest}) \over d{\phi}} = {C{\eta} \over C + {\eta}}{dV_{rest} \over d{\phi}} $$
or on using Eq. 13
\begin{equation}
dQ_i = {C{\eta} \over C + {\eta}}{dV_{rest}} 
\end{equation}
or
\begin{equation}
dQ_i = -dQ_{rest}=C_{eff}dV_{rest} 
\end{equation}
where, $ C_{eff} = {C{\eta} \over C + {\eta}} $
That is
\begin{equation}
{1 \over C_{eff}}={1 \over C} + {1 \over \eta}
\end{equation}
When we define capacitance we do not consider the sign of the charge.
Normally one plate of the capacitor has charge $+Q$ and the other has
charge $-Q$, wherein we write $Q=CV$.
Hence if we want to include quantum effects then only in the semi-classical
regime we find it possible to describe polarization in terms of
capacitance.

A quantum capacitor of capacitance $\eta$ in series with the classical
capacitance determines the effective capacitance of the system.
The characteristic potential (or the potential difference between
two parts of the ring) is determined by this effective variable.
The AC response of the ring is also determined by this effective
variable along with the inductance. For a purely capacitive response,
$I(t)=dQ_i/dt$ or $dI_\omega =-i\omega C_{eff}dV_\omega$
\cite{scr}.
Such an effective variable will exist only if assumptions given
in Eq. 3, Eq. 8, and Eq. 15 are valid.
Although quantum capacitance was introduced first by Serge Luryi \cite{ser},
he introduced it on very general grounds and the above
relation was not obtained. The above relation was obtained
for a closed system
only in the frame work of single particle two level system \cite{scr}.
We have derived it generally for any arbitrary system including
many body effects and outlined the assumptions so that
such an effective variable will exist.

In Figs. 6 and 7 we have made a comparison of quantum capacitance
at site 8, i.e., $\eta_8$ in the
non-interacting system and the interacting system to show that
$\eta$ is a good parameter to effectively capture the effect
of Coulomb repulsion. Here again we are considering the
11 site disordered ring considered in Figs. 3 and 4.
We have applied a small external potential at the 8th site
to evaluate $\partial Q_8/\partial V_8=\eta_8$. 
Fig. 6 is for the non-interacting
case i.e., $U_1=0$ and $U_2=0$ although the electrons still interact
through Fermi statistics. In Fig 7 we have made $U_1=2$ with all
other parameters remaining the same. There is a large
qualitative as well as quantitative difference between the
two figures which shows the importance of including Coulomb
interaction and many body effects in defining capacitance.
$dQ_i=-dQ_{rest}$ can be measured as outlined in Fig. 2.
${\partial Q_i \over \partial V_i} =-\eta$ can also be measured
as outlined in Fig. 2. 
Geometrical capacitance $C$ is independent of finite size or
quantum interference effects and is known for a given sample
from its bulk properties.
So one can determine $C_{eff}$. 
Thus $dV_i$ and $dV_{rest}$ can be known
for any applied external potentials as $C_{eff}$ is the single
parameter that determines this.

\section{conclusion}

In conclusion, in the semi-classical regime we prove
polarization charge and induced potential
of a mesoscopic isolated sample are related by an
effective capacitance $C_{eff}$.
Effective capacitance can be decoupled
as a linear combination of classical capacitance and quantum
capacitance. The quantum capacitance is given by the Lindhard
function. In this regime, we can design quantum circuits in
terms of this parameter $C_{eff}$ just as classical circuits
are built in terms of parameters like resistance, capacitance
and inductance.
While in earlier works , Eq. 18 was derived for single particle theory for a
two level system, we have started from the principles of many
body physics and derived Eq. 18 for any arbitrary geometry
in any dimension.
We have shown the approximations necessary to get this in the
framework of fully interacting fermions. 
All these approximations are
justified in the semi-classical limit. 
So Eq. 18 can provide a simple way to understand
DC and AC response of quantum finite sized many body electronic
systems in the semi-classical limit, in terms of an effective variable.

The authors acknowledge useful discussions with Dr. P. Koskinen
and Dr. S. Gupta. One of us (PSD) also thanks DST for funding this
research.

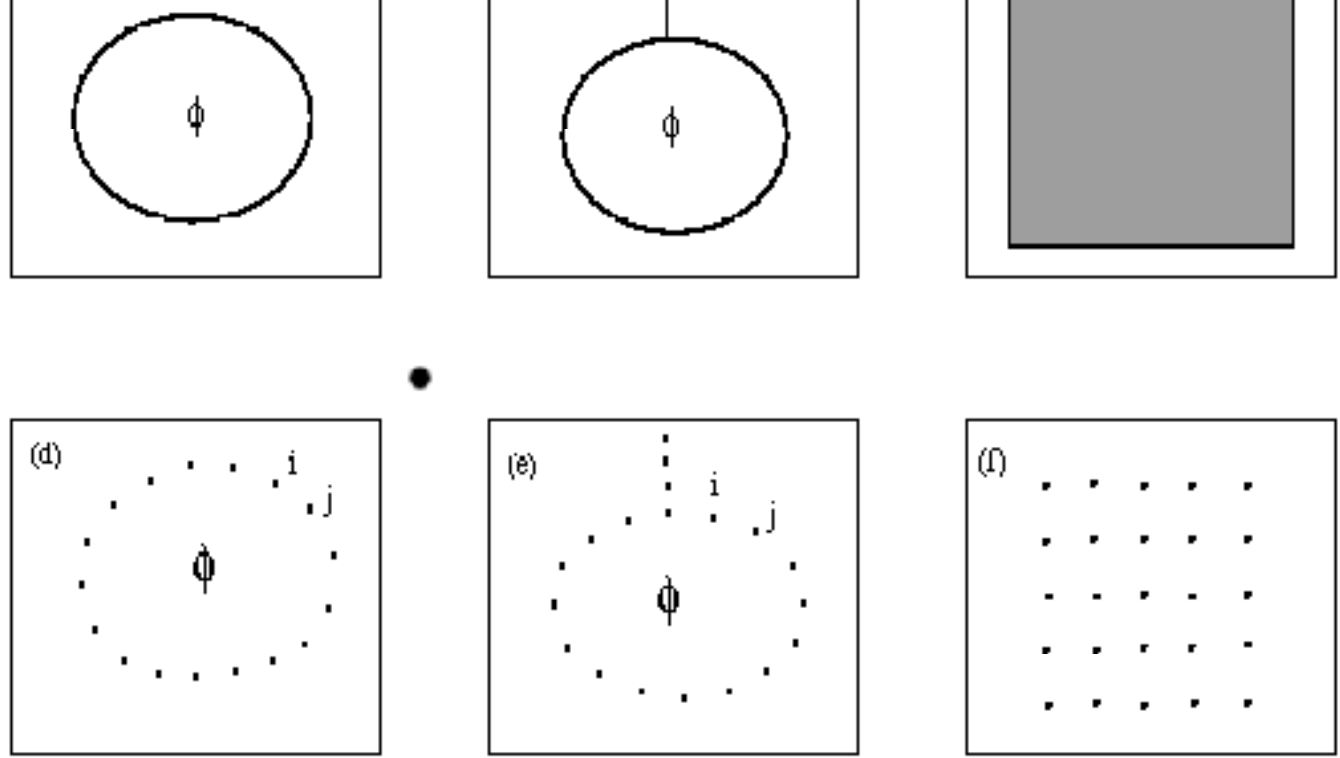
Fig. 1. Schematic diagrams of some mesoscopic geometries used in this 
work as examples and also for numerical verifications. Fig. 1(a)
represents a one-dimensional ring pierced by a magnetic flux $\phi$.
Fig. 1(b) represents a one-dimensional ring to which a quantum wire
or stub is attached. The ring is again pierced by a magnetic flux $\phi$.
Fig. 1(c) represents a
two-dimensional square geometry. Once again a magnetic field
can be applied perpendicular to the plane of the geometry.
Figs. 1(d), 1(e) and 1(f) are discrete versions of those in
1(a), 1(b) and 1(c) respectively, that can be described by
a generalized Hubbard Hamiltonian and useful for numerical
verifications. The dots represent sites. Nearest neighbor
sites are marked $i$ and $j$.

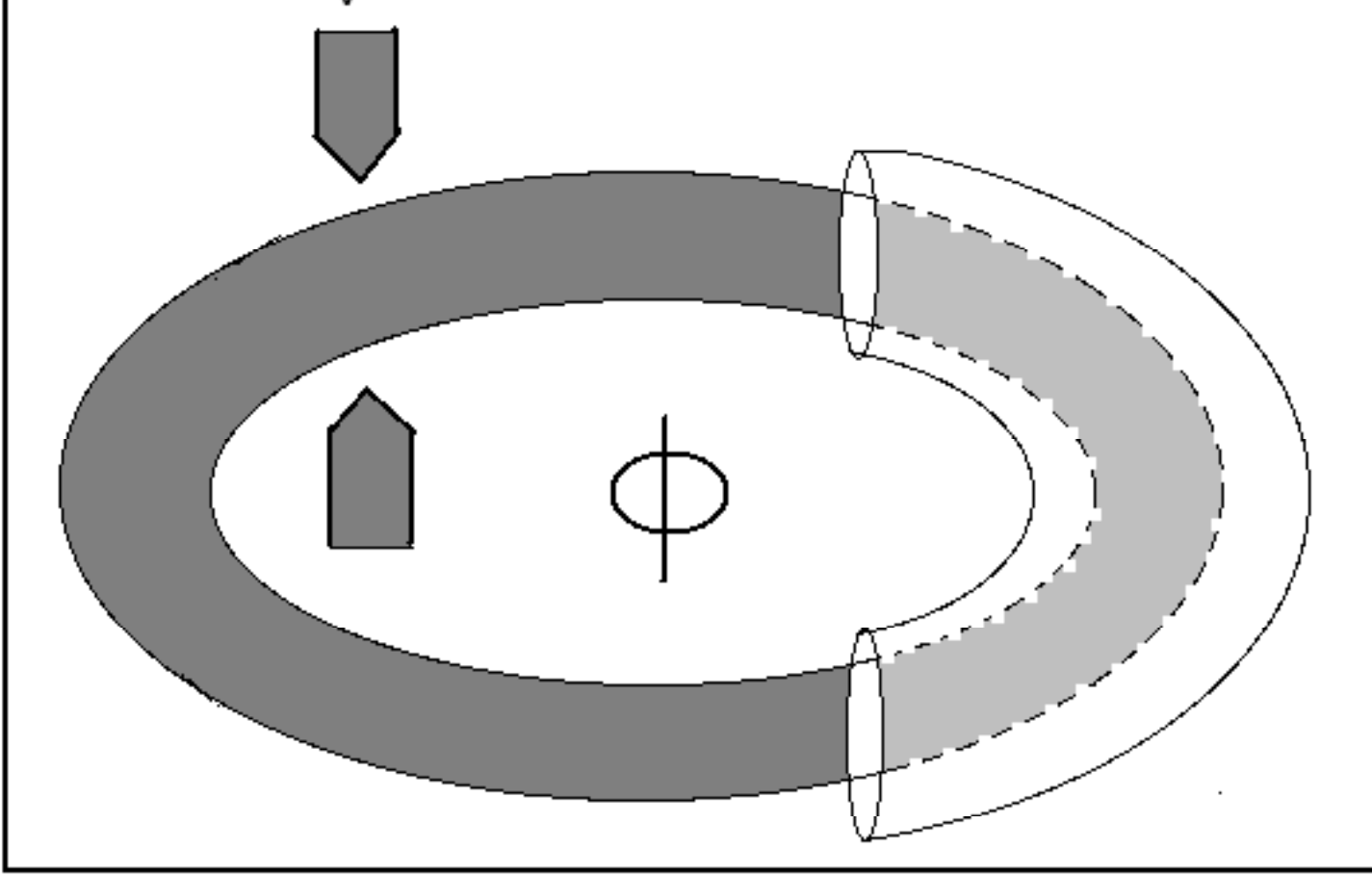
Fig. 2. A three-dimensional mesoscopic ring pierced by a magnetic
flux $\phi$. The ring can 
be polarized by a voltage probe (without making a contact) at
site $r$ at a given voltage $V$.
The flux can also polarize the ring. Another voltage probe (without making
contact) whose voltage is allowed to vary can measure the voltage
at the site $r$. A solenoid around the ring, as shown in the figure,
can measure the induced charge in a segment of the ring due to 
polarization. The idea of quantum capacitance is valid only when
the polarized charge is uniformly distributed in the rest of the ring,
apart from the region at $r$.

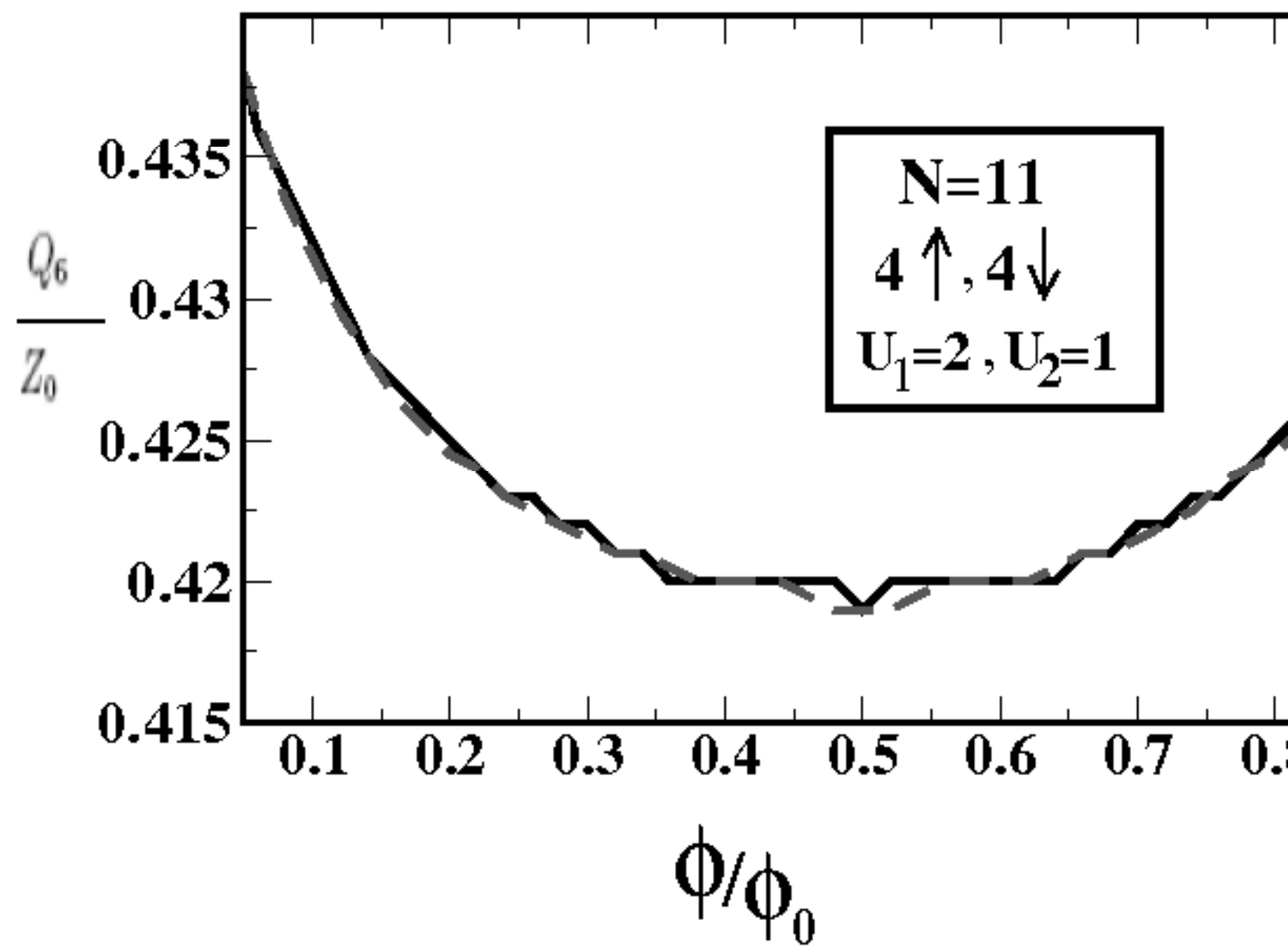
Fig. 3. The figure shows that for the geometries shown in Fig. 1,
Eq. 3 is valid. In this figure we have used the geometry in
Fig. 1(d). It consists of 11 sites ($N=11$), with 4 spin up electrons
and 4 spin down electrons. The on-site Hubbard $U_1=2$, the nearest
neighbor Hubbard $U_2=1$. The hopping parameter $t=1$.
The solid line is the charge on the 6th site as a function of the
flux in units of $Z_0$ which is just electronic charge taken to be 1. 
The dashed line is ${\partial E \over \partial V_6}$ in units of
$Z_0=$ electronic charge taken as 1. 
$E$ is the ground state eigen-energy of the many body 
system found by exact diagonalization using Lanczos algorithm.
The dimension of the matrix being of the order $10^5X10^5$.
Here $\phi_0=hc/e$. We have also verified Eq. 3 for the geometries
in Fig. 1(e) and 1(f).

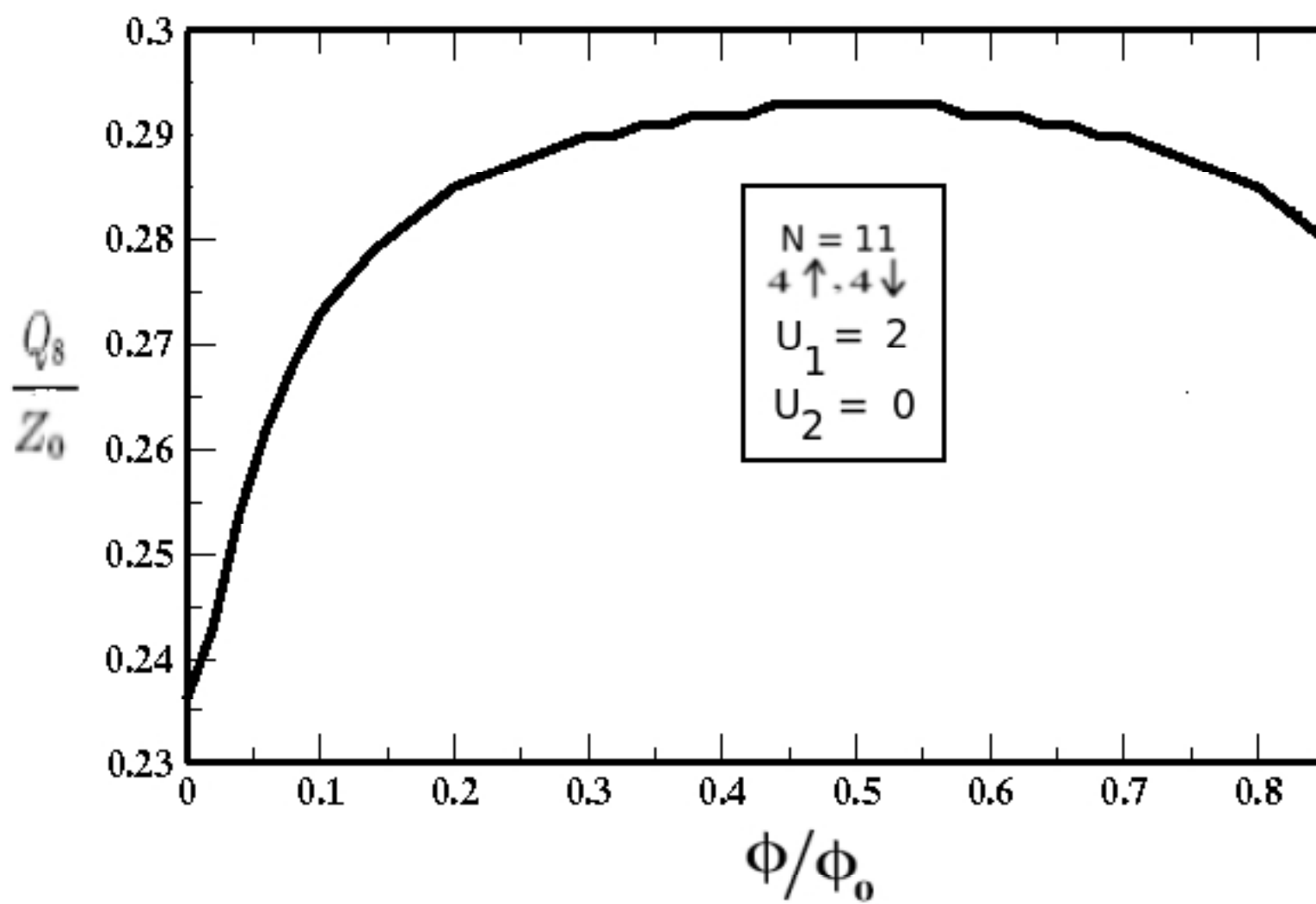
Fig. 4. The fig shows that a disordered
mesoscopic ring can be polarized by
an Aharonov-Bohm flux alone. We have used the geometry of Fig 1(d)
here. Parameters used are shown in the inset. We have used only
one disorder configuration for this figure with site energies
varying from $-.5t$ to $.5t$, but we have also checked
for other disorder configurations.
The graphs for other geometries is qualitatively similar and so
not shown here. $Q_8$ is the charge density at the 8th site, in
units of $Z_0$ which is electronic charge. At zero flux we expect
the system to be neutral. As the flux changes, strong dispersion
of $Q_8$ suggests polarization of the system wherein the
positive charge in the system can be taken to be uniform and
independent of flux.
Here $\phi_0=hc/e$.

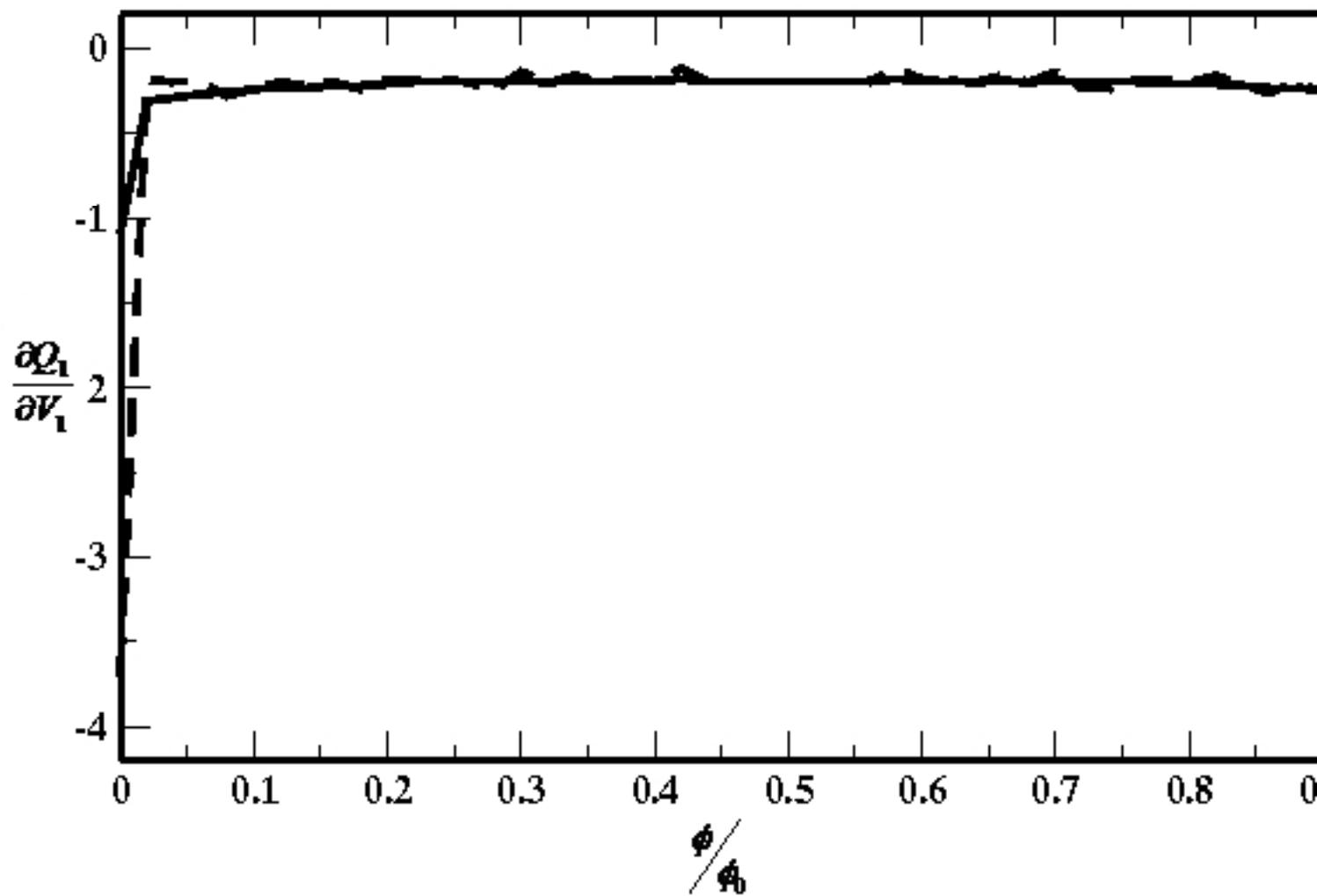
Fig. 5. Here we are considering a ring in the semi classical limit.
That is the potential in the ring varies very slowly compared to
de-Broglie wavelength. In particular we have taken a 11 site ring with
a single defect, $\epsilon_1=0.3$ and rest of the site energies
are 0. The solid line is $\partial Q_1 \over \partial V_1$ and
the dashed line is -$\sum_{j\ne 1}{\partial Q_1 \over \partial V_j}$.
Both quantities are in units of electronic charge taken as 1.
Here $U_1=2$ and $U_2=1$ with 4 spin up and 4 spin down electrons
in the ring.
Here $\phi_0=hc/e$.

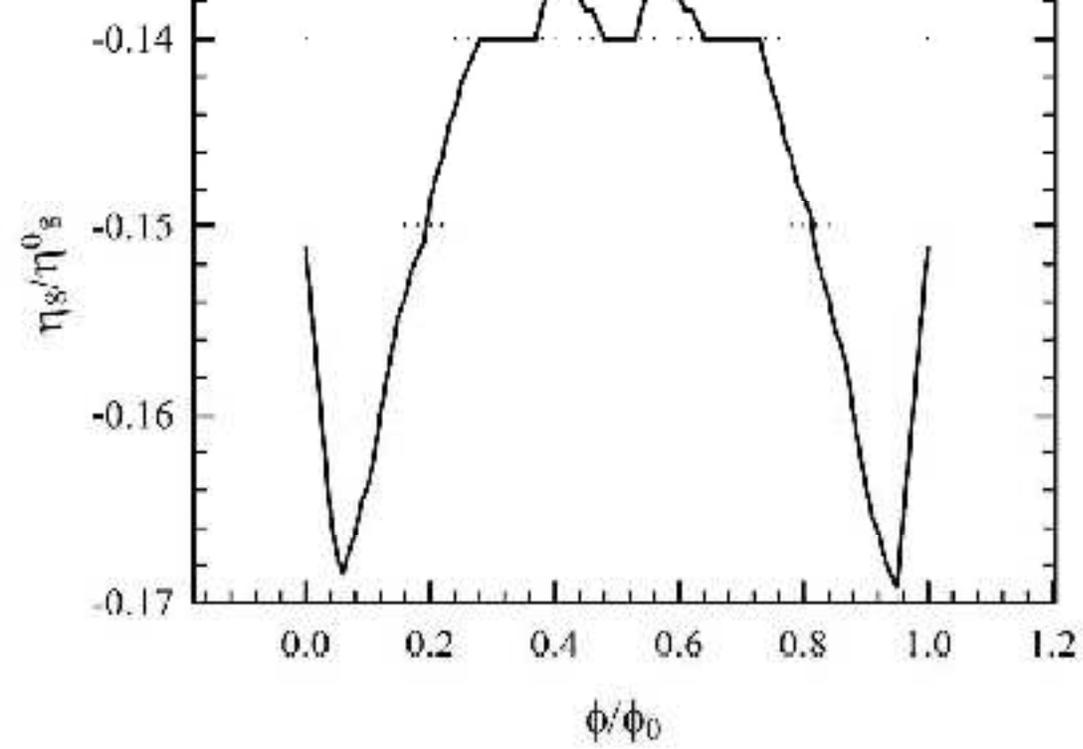
Fig. 6. The figure shows a plot of $\eta_8=\partial Q_8/\partial 
V_8$ as a function of
flux. Here $\eta^0_8=e/t$ where $e$ is electronic charge. $\phi_0=hc/e$.
This is for an eleven site disordered ring with site energies varying
from $-.5t$ to $+.5t$. In this case $U_1=0$ and $U_2=0$.
Which means electrons are interacting only through Fermi statistics.
Here $\phi_0=hc/e$.

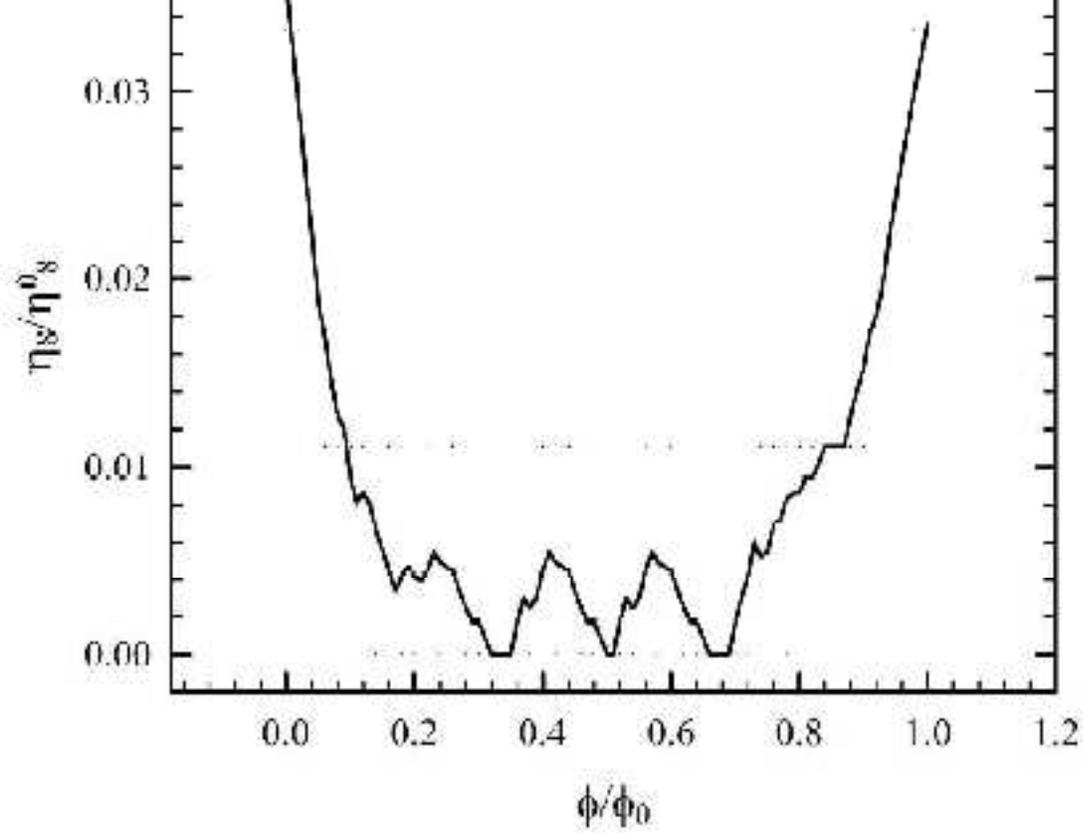
Fig. 7. The figure shows a plot of $\eta_8=\partial Q_8/\partial 
V_8$ as a function of
flux. Here $\eta^0_8=e/t$ where $e$ is electronic charge. $\phi_0=hc/e$.
This is for an eleven site disordered ring with site energies varying
from $-.5t$ to $+.5t$. In this case $U_1=2$ and $U_2=0$.
Here $\phi_0=hc/e$.

\end{document}